# Degree of Separation in Social Networks

Prerana Laddha

***ABSTRACT***

According to the small-world concept, the entire world is connected through short chains of acquaintances. In popular imagination this is captured in the phrase *six degrees of separation*, implying that any two individuals are, at most, six handshakes away. Social network analysis is the understanding of concepts and information on relationships among interacting units in an ecological system. In this analysis the properties of the actors are explained in terms of the structures of links amongst them. In general, the relational links between the actors are primary and the properties of the actors are secondary. This paper presents two methods to calculate the average degree of separation between the actors or nodes in a graph. We apply this approach to other random graphs depicting social networks and then compare the characteristics of these graphs with the average degree of separation.

***KEYWORDS***

Nodes, Links, Degree of separation, Probability of connectivity.

## I INTRODUCTION

The small world phenomenon -- the proposition that says that the world is connected by a *six* degree of separation -- was demonstrated by the social psychologist Stanley Milgram in the year 1967[1]. Milgram carried out his experiment by asking various people in different cities to send a letter to specifically named targets that were not known to the senders. The condition placed on each sender was that he or she could mail the letter to someone with whom the sender was on the first name basis and to someone who is most likely to forward the letter to its final destination. The letters included information, which explained the purpose of the experiment, and basic information about the target contact person. By observing the results from the different series of letters that reached the target, Milgram concluded that the average degree of separation was *six*. This was the origin of the six degree of separation theory.

Since social networks may be viewed through the ecological point of view, the idea of low degree of separation must be true of other ecological networks. The brain itself is an ecological system [2],[3],[4] and its structural expression is in terms of recursive structure, which makes the separation between components small [5],[6],[7],[8]. This structure is hierarchical in neural network models [9],[10],[11] of memory. Brain models, therefore, have aspects that are similar to social networks, whether they are small world networks or scale-free networks. A study of degree of separation in social networks can eventually have applications in other ecological networks.

In this paper we represent the networks in a sequentially arranged tree-like graph structure and present two mathematical methods using matrix operations to calculate the average degree of separation between the nodes in the graph. Graph theory also serves as a utility to represent the social network as a model of the social environment, in which the actors represent the nodes and relationship between them is the edges in the graph. We extend this concept by applying the second method to random graphs.

Random graphs generated from the Watts and Strogatz model [12] that depict the small world in a convenient manner are considered and the average degree of separation is found. We also observe the relationship between the probability of connectivity between the nodes and the average degree of separation. A network may also be represented by statistical and probability methods and algebraic models [13].

The rest of the paper is organized into three sections. Section II contains the two different approaches used to find the degree of separation between the nodes. In section III the second technique is applied on random graphs and we draw certain conclusions on the connectivity property of the graphs with respect to the degree of separation. Section IV concludes the paper.

## II AVERAGE DEGREE OF SEPARATION

Amongst the various possible ways in which a network can be viewed, one of the most useful views is as a graph. In this paper, a sequentially increasing graph is considered to depict the network structure. This graphical representation can be adapted to represent a wide range of network data. The nodes are arranged such that they form a tree like structure as shown in the figure1. This format of network is chosen over others because of its clarity and efficiency and also it helps the readers to increase their understandability about the average degree of separation in a network. In this paper, the average degree of separation of the nodes is calculated using two different methods.

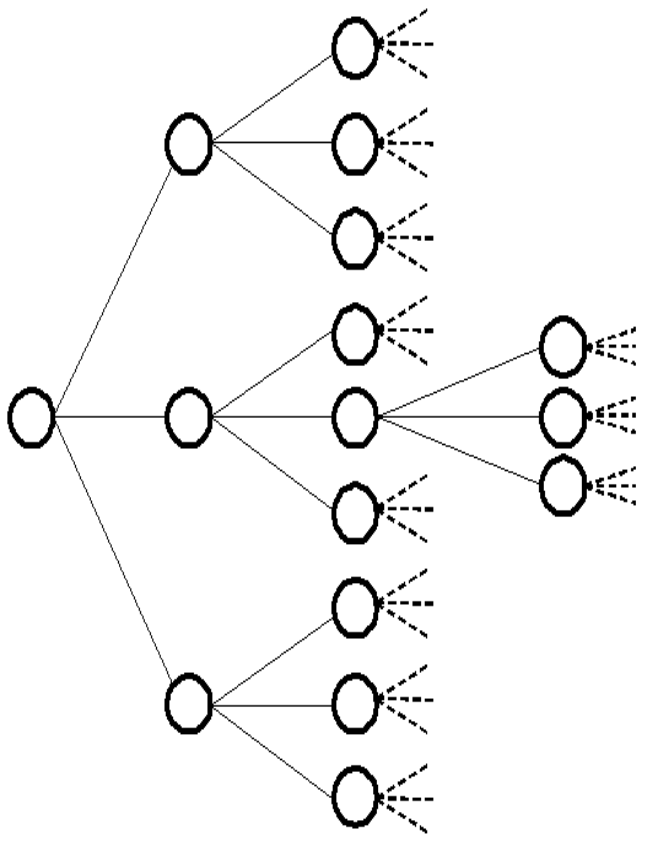

Figure 1: Tree showing the sequential arrangement of nodes

In the figure 1 let us consider that we have a set of actors, which will be denote by $N = \{n_1, n_2, n_3 \ldots\ldots n_k\}$. Suppose that node $n_1$ has $r$ links that are associated with it. Taking one step further, consider that each of these $r$ links have another set of $r$ links. This results in a total number of $r^2 + r + 1$ links approximately. As the process continues, we get $r^3$ number of links in third step and similarly $r^k$ number of links in $k^{th}$ step. Figure 1 gives the graphical representation of this process. In this way we can arrange any number of nodes for a given degree $r$. This gives a geometric series as follows.

$$r^0 + r^1 + r^2 + r^3 \ldots\ldots r^k = N$$

Here $N$ is the total number of nodes that are being considered. The geometric series in the above equation can be represented as below.

$$(1- r^k) / (1-r) = N$$

In this paper, two methods are proposed to calculate the average degree of separation in the sequential graph. In the first method, a table is created in order to calculate the number of steps taken to reach all possible nodes from each level in the graph. The rows in the table are indexed as $S_1$, $S_2 \ldots S_k$, each of them indicating a particular level in the tree. $S_1$ depicts the first level, consisting of $r^0$ nodes. $S_2$ depicts the second level as the depth of the tree increases by one. The numbers of nodes at this level are $r^1$. Similarly the numbers of nodes in the $S_k$ stage are $r^k$. This table can be generalized for a particular value of $k$.

The tables in the figure 2 shows the generalized formats for $k$=3, 4, 5 and 6. By substituting any value of $r$ in any of these tables for a given value of $k$, we can obtain the required values. For example, if the given values are $r=3$ and $k=4$, we substitute the value of r in the second table i.e. Table 2, in order to get the required values. Each value in the table gives the number of nodes that can be reached from that particular level in the given number of steps. The values that are obtained in the table are then written in the form of matrix which is of the order $k$ x $l$. Here $l$ is the maximum number of steps between any two nodes in the graph. This value is obtained by applying the formula $l = 2(k-1)$ .

In order to calculate the average degree of separation between the nodes in a graph, we consider one node in each of the steps, calculate the average and multiply it with the total number of nodes in that step. Initially, we consider a specific node in $S_k$ and count the number of nodes that can be reached by that node in $1, 2, 3\ldots l$ number of steps. This number is multiplied with the respective degree of separation and then the total is divided by the total nodes in the graph excluding itself. This gives the average degree of separation of any node that belongs to step $S_k$. In this way one node is selected from each step and the degree of separation is calculated. This process is followed by the remaining nodes in the graph. This gives the average degree of separation of each node in each step of the graph. Finally, number of nodes in each step is multiplied by its respective average in order to get the total average degree of separation of that graph.

Thus, the average degree of separation is given by the following formula. Here $i$ represents the rows and $j$ represents the columns in the matrix. $l$ here gives the number of columns in the matrix , $k$ is the number of steps in the sequential graph and $N$ is the number of nodes in the graph.

$$\sum_{i=1}^{k} \frac{[\sum_{j=1}^{l}(x_{ij} * j)/N-1] * r^{k-i}}{N}$$

Figure 2: The following tables give the degree of separation from each step in the tree. The steps are denoted as $S_1$, $S_2$, $S_3$... $S_k$. The columns are indexed with the numbers 1, 2, 3 and so on denoting the number of steps the node takes in that level to reach other nodes.

| | 1 | 2 | 3 | 4 |
|---|---|---|---|---|
| $S_3$ | $r^0$ | $r^1$ | $r-1$ | $(r-1)r$ |
| $S_2$ | $r+1$ | $r-1$ | $(r-1)r$ | - |
| $S_1$ | $r^1$ | $r^2$ | - | - |

Table 1: Generalized values of r for k=3

| | 1 | 2 | 3 | 4 | 5 | 6 |
|---|---|---|---|---|---|---|
| $S_4$ | $r^0$ | $r^1$ | $r^1$ | $r^2-1$ | $r^2-r$ | $r^3-r^2$ |
| $S_3$ | $r+1$ | $r^1$ | $r^2-1$ | $r^2-r$ | $r^3-r^2$ | - |
| $S_2$ | $r+1$ | $r^2+r-1$ | $r^2-r$ | $r^3-r^2$ | - | - |
| $S_1$ | $r^1$ | $r^2$ | $r^3$ | - | - | - |

Table 2: Generalized values of r for k=4

| | **1** | **2** | **3** | **4** | **5** | **6** | **7** | **8** |
|---|---|---|---|---|---|---|---|---|
| $S_5$ | $r^0$ | $r^1$ | $r^1$ | $r^2$ | $r^2$-1 | $(r^2-1)r$ | $r^3-r^2$ | $r^4-r^3$ |
| $S_4$ | $r^1+1$ | $r^1$ | $r^2$ | $r^2-1$ | $(r^2-1)r$ | $r^3-r^2$ | $r^4-r^3$ | - |
| $S_3$ | $r^1+1$ | $r^1+r^2$ | $r^2-1$ | $(r^2-1)r$ | $r^3-r^2$ | $r^4-r^3$ | - | - |
| $S_2$ | $r^1+1$ | $r^1+r^2-1$ | $(r^1+r^2-1)r$ | $r^3-r^2$ | $r^4-r^3$ | - | - | - |
| $S_1$ | $r^1$ | $r^2$ | $r^3$ | $r^4$ | - | - | - | - |

Table 3: Generalized values of r for k=5

| | **1** | **2** | **3** | **4** | **5** | **6** | **7** | **8** | **9** | **10** |
|---|---|---|---|---|---|---|---|---|---|---|
| $S_6$ | $r^0$ | $r^1$ | $r^1$ | $r^2$ | $r^2$ | $r^3-1$ | $r^1+r^2$ | $r^2+r^3$ | $r^4-r^3$ | $r^5-r^4$ |
| $S_5$ | $r^1+1$ | $r^1$ | $r^2$ | $r^2$ | $r^3-1$ | $r^1+r^2$ | $r^2+r^3$ | $r^4-r^3$ | $r^5-r^4$ | - |
| $S_4$ | $r^1+1$ | $r^1+r^2$ | $r^2$ | $r^3-1$ | $r^1+r^2$ | $r^2+r^3$ | $r^4-r^3$ | $r^5-r^4$ | - | - |
| $S_3$ | $r^1+1$ | $r^1+r^2$ | $r^3+r^2-1$ | $r^1+r^2$ | $r^2+r^3$ | $r^4-r^3$ | $r^5-r^4$ | - | - | - |
| $S_2$ | $r^1+1$ | $r^2+r^1-1$ | $(r^2+r^1-1)r$ | $(r^3+r^2-1)r^2$ | $r^4-r^3$ | $r^5-r^4$ | - | - | - | - |
| $S_1$ | $r^1$ | $r^2$ | $r^3$ | $r^4$ | $r^5$ | - | - | - | - | - |

Table 4: Generalized values of r for k=6

**Example 1.** A small graph of forty nodes is considered and a mathematical calculation for the average number of steps for two nodes to communicate with each other is made. In this process, we consider that the forty nodes are arranged in the form of a sequential tree, where each node is further attached to three other nodes. Figure 3 shows the diagrammatic representation of the nodes in the form of a tree. From the figure we can observe that there are four sequentially increasing steps in the graph, which gives us the value of $k$ as 4. We name the steps as $S_1$, $S_2$, $S_3$ and $S_4$. The number of nodes in $S_1$ are *1* i.e. $3^0$, in $S_2$ there are *3* i.e. $3^1$ nodes and so on. Therefore the total number of nodes in the tree is given by a geometric series which is as follows.

$$3^0+3^1+3^2+3^3=40$$

In this example the values are $r=3$, $k=4$, $N=40$, $l=6$. By substituting the values of $r=3$ in table 2 from the figure 2, we get the following values.

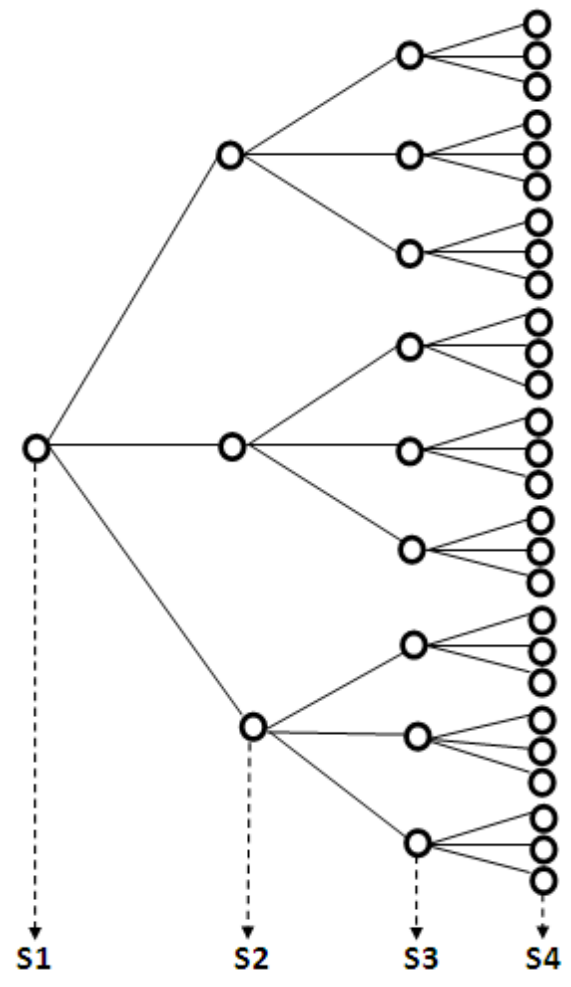

Figure 3: Tree arrangement of 40 nodes

| | 1 | 2 | 3 | 4 | 5 | 6 |
|---|---|---|---|---|---|---|
| $S_4$ | 1 | 3 | 3 | 8 | 6 | 18 |
| $S_3$ | 4 | 3 | 8 | 6 | 18 | - |
| $S_2$ | 4 | 11 | 6 | 18 | - | - |
| $S_1$ | 3 | 9 | 27 | - | - | - |

$$\begin{bmatrix} 1 & 3 & 3 & 8 & 6 & 18 \\ 4 & 3 & 8 & 6 & 18 & - \\ 4 & 11 & 6 & 18 & - & - \\ 3 & 9 & 27 & - & - & - \end{bmatrix}$$

Table 5: The values obtained for *r=3* and *k=4* from figure 2, written in the form of a matrix

The values in the table are written in the form of a matrix of order 4x6 and by substituting the values in the equation we get the average degree of separation as *4.233*.

In the second method, the information in the graph is expressed in the matrix form. In this method we calculate the average degree of separation by making use of a matrix that consists of all the information that is related to the nodes and links of the sequential graph. Each value in the matrix gives the shortest distance between the two nodes. The matrix thus obtained will also be symmetric. In order to obtain this matrix we use the algorithm below:

Step 1: Consider a $N$ x $N$ matrix and initially, $\forall\ i = j\ , x_{ij} = 0$.

Step 2: Write an adjacency matrix or a sociomatrix for the given graph.

Step 3: For $p=1, 2, 3 \ldots l\text{-}1$, $\forall\ x_{ij} = p$ in the $i$th row, find all $x_{jk} = 1$ in the $j$th row and write $x_{ik} = p + 1,\ \ iff\ \ x_{ik} = 0\ or\ x_{ik} > p + 1$

Step 4: Repeat the process till the all the values in the matrix are filled.

Step 5: Calculate $sum = \sum_{i=1}^{N} \sum_{j=1}^{N} x_{ij}$

Step 6: Calculate Average degree of separation = $sum / (N-1)^2$

Figure 4: Tree arrangement of 15 nodes

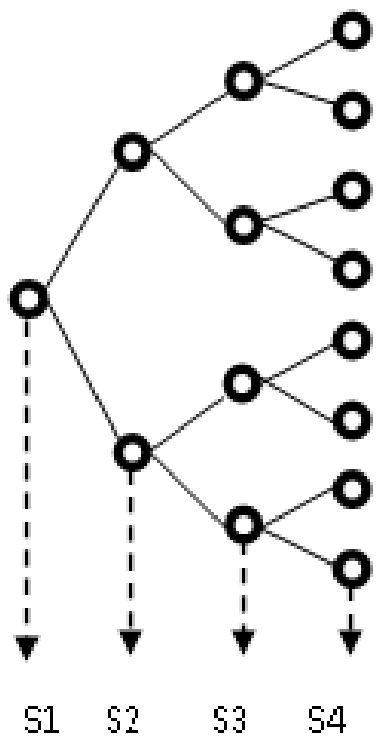

**Example 2.** In this example we arrange fifteen nodes in the form of a sequential tree. Here the nodes are attached to two other nodes in a sequentially increasing manner and are arranged in four steps. The geometric series associated with this kind of tree structure is given below

$2^0 + 2^1 + 2^2 + 2^3 = 15$

The values are *r=2, k=4, N=15, l=6.* The matrix obtained by following the above algorithm is as follows

Adjacency matrix

$$\begin{bmatrix}
0&1&1&0&0&0&0&0&0&0&0&0&0&0&0\\
1&0&0&1&1&0&0&0&0&0&0&0&0&0&0\\
1&0&0&0&0&1&1&0&0&0&0&0&0&0&0\\
0&1&0&0&0&0&0&1&1&0&0&0&0&0&0\\
0&1&0&0&0&0&0&0&0&1&1&0&0&0&0\\
0&0&1&0&0&0&0&0&0&0&0&1&1&0&0\\
0&0&1&0&0&0&0&0&0&0&0&0&0&1&1\\
0&0&0&1&0&0&0&0&0&0&0&0&0&0&0\\
0&0&0&1&0&0&0&0&0&0&0&0&0&0&0\\
0&0&0&0&1&0&0&0&0&0&0&0&0&0&0\\
0&0&0&0&1&0&0&0&0&0&0&0&0&0&0\\
0&0&0&0&0&1&0&0&0&0&0&0&0&0&0\\
0&0&0&0&0&1&0&0&0&0&0&0&0&0&0\\
0&0&0&0&0&0&1&0&0&0&0&0&0&0&0\\
0&0&0&0&0&0&1&0&0&0&0&0&0&0&0
\end{bmatrix} \Rightarrow$$

Matrix obtained when $p$=1

$$\begin{bmatrix}
0&1&1&2&2&2&2&0&0&0&0&0&0&0&0\\
1&0&2&1&1&0&0&2&2&2&2&0&0&0&0\\
1&2&0&0&0&1&1&0&0&0&0&2&2&2&2\\
2&1&0&0&2&0&0&1&1&0&0&0&0&0&0\\
2&1&0&2&0&0&0&0&0&1&1&0&0&0&0\\
2&0&1&0&0&0&2&0&0&0&0&1&1&0&0\\
2&0&1&0&0&2&0&0&0&0&0&0&0&1&1\\
0&2&0&1&0&0&0&0&2&0&0&0&0&0&0\\
0&2&0&1&0&0&0&2&0&0&0&0&0&0&0\\
0&2&0&0&1&0&0&0&0&0&2&0&0&0&0\\
0&2&0&0&1&0&0&0&0&2&0&0&0&0&0\\
0&0&2&0&0&1&0&0&0&0&0&0&2&0&0\\
0&0&2&0&0&1&0&0&0&0&0&2&0&0&0\\
0&0&2&0&0&0&1&0&0&0&0&0&0&0&2\\
0&0&2&0&0&0&1&0&0&0&0&0&0&2&0
\end{bmatrix}$$

Matrix obtained when $p$=2

$$\begin{bmatrix}
0&1&1&2&2&2&2&3&3&3&3&3&3&3&3\\
1&0&2&1&1&3&3&2&2&2&2&0&0&0&0\\
1&2&0&3&3&1&1&0&0&0&0&2&2&2&2\\
2&1&3&0&2&0&0&1&1&3&3&0&0&0&0\\
2&1&3&2&0&0&0&3&3&1&1&0&0&0&0\\
2&3&1&0&0&0&2&0&0&0&0&1&1&3&3\\
2&3&1&0&0&2&0&0&0&0&0&3&3&1&1\\
3&2&0&1&3&0&0&0&2&0&0&0&0&0&0\\
3&2&0&1&3&0&0&2&0&0&0&0&0&0&0\\
3&2&0&3&1&0&0&0&0&0&2&0&0&0&0\\
3&2&0&3&1&0&0&0&0&2&0&0&0&0&0\\
3&0&2&0&0&1&3&0&0&0&0&0&2&0&0\\
3&0&2&0&0&1&3&0&0&0&0&2&0&0&0\\
3&0&2&0&0&3&1&0&0&0&0&0&0&0&2\\
3&0&2&0&0&3&1&0&0&0&0&0&0&2&0
\end{bmatrix} \Rightarrow \ldots\ldots$$

Matrix obtained when $p=l-1$

$$\begin{bmatrix}
0&1&1&2&2&2&2&3&3&3&3&3&3&3&3\\
1&0&2&1&1&3&3&2&2&2&2&4&4&4&4\\
1&2&0&3&3&1&1&4&4&4&4&2&2&2&2\\
2&1&3&0&2&4&4&1&1&3&3&5&5&5&5\\
2&1&3&2&0&4&4&3&3&1&1&5&5&5&5\\
2&3&1&4&4&0&2&5&5&5&5&1&1&3&3\\
2&3&1&4&4&2&0&5&5&5&5&3&3&1&1\\
3&2&4&1&3&5&5&0&2&4&4&6&6&6&6\\
3&2&4&1&3&5&5&2&0&4&4&6&6&6&6\\
3&2&4&3&1&5&5&4&4&0&2&6&6&6&6\\
3&2&4&3&1&5&5&4&4&2&0&6&6&6&6\\
3&4&2&5&5&1&3&6&6&6&6&0&2&4&4\\
3&4&2&5&5&1&3&6&6&6&6&2&0&4&4\\
3&4&2&5&5&3&1&6&6&6&6&4&4&0&2\\
3&4&2&5&5&3&1&6&6&6&6&4&4&2&0
\end{bmatrix}$$

Table 1: Example of a 15 x 15 matrix that is obtained by following the above algorithm for the graph in figure 4.

By following the steps 5 and 6 in the algorithm, we get the value of average degree of separation of the graph as *3.74*.

**Example 3.** We consider a structured graph which is connected in a regular fashion [3]. Here the graph is symmetric and every node is connected to $2\log_2(N)-1$ other nodes in a regular fashion. The nodes are connected such that, each node connects to *i*+1, *i*+2, *i*+4, *i*+8, *i*+16… *i*+$\log_2(\frac{N}{2})$ other nodes [3]. In figure 5, for $N$ = 16 the symmetric connections of the nodes are depicted. We apply the second method to it using a 16x16 adjacency matrix and after simulation the average degree of separation is found to be 1.63. The theorem 1 stated in the key distribution scheme for sensor networks [3] also holds good. This can be verified by comparing the greatest value in the matrix to $\left\lfloor \frac{\log_2(\frac{N}{8})}{2} \right\rfloor + 2$. Here this value equals to 2.

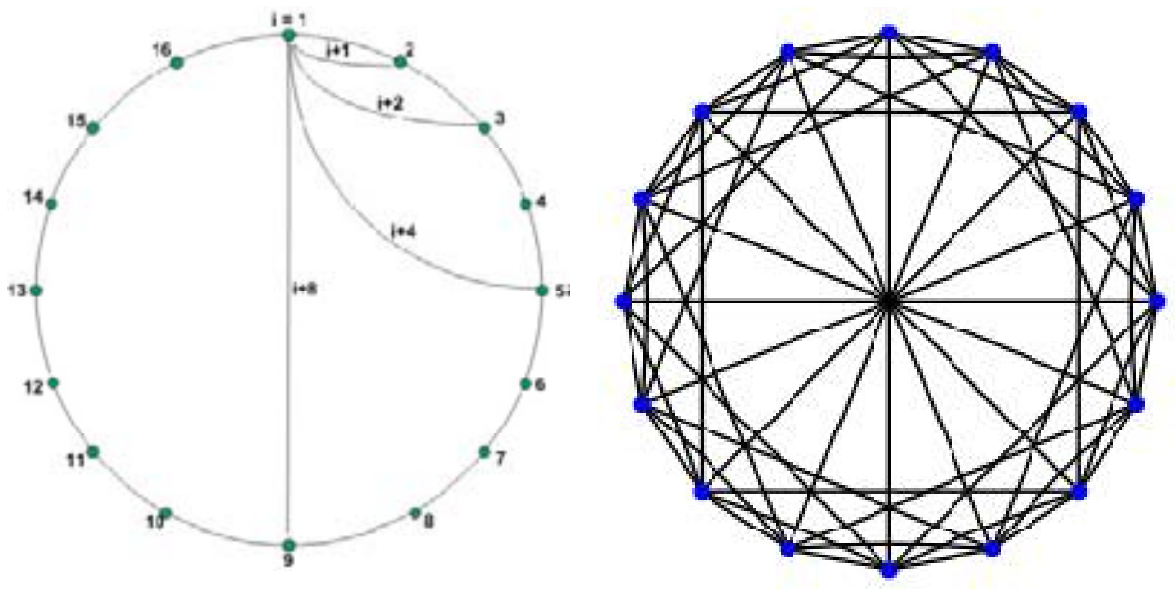

Figure 5: a structured symmetric graph , that is connected in a regular fashion[3]

### III RANDOM GRAPHS

Social networking has provided impetus for development of various network models and methods. Researchers have started to find the use of mathematical models for social network analysis. Amongst all the various methods that have been researched by far, the Watts and Strogatz model is considered as the simplest model that depicts various small world properties [2]. Here, we apply the matrix method to find the degree of separation in a graph with *20* nodes. We also relate the average degree of separation to the changing probability in the graph. The following figure depicts the change in the connectivity of a graph according to watts and strogatz model.

Figure 6: A graph consisting of 20 nodes which is rewired with different probability of connectivity according to watts and strogatz theory[2]

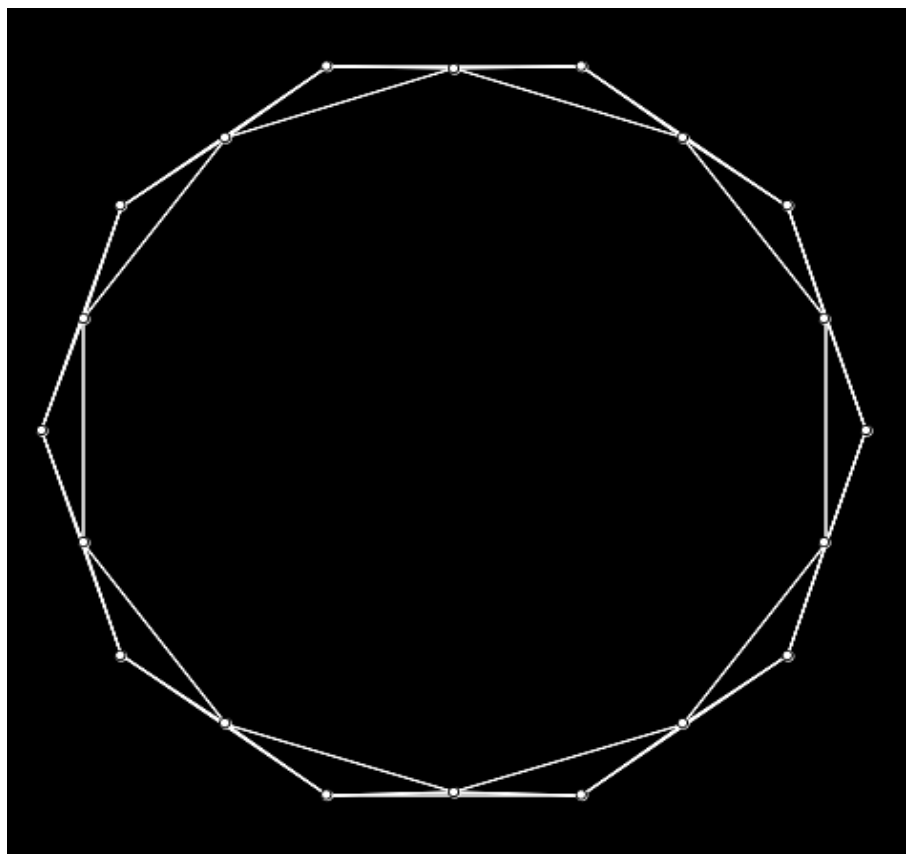

a) Probability = 0.00

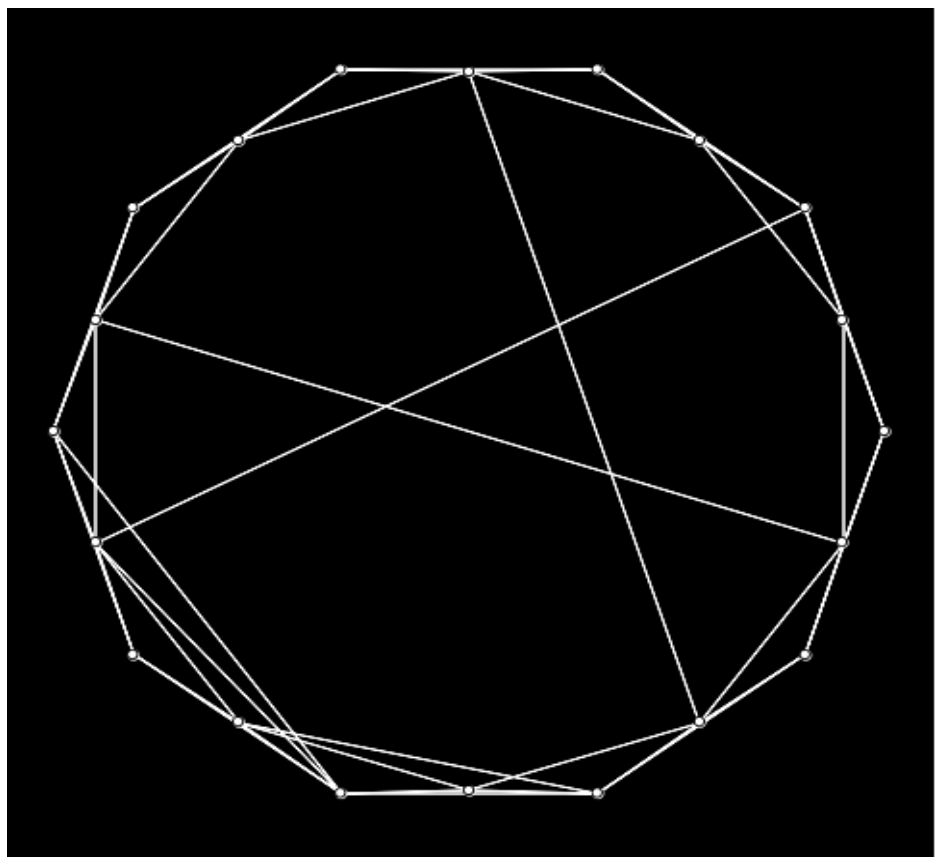

b) Probability =0.20

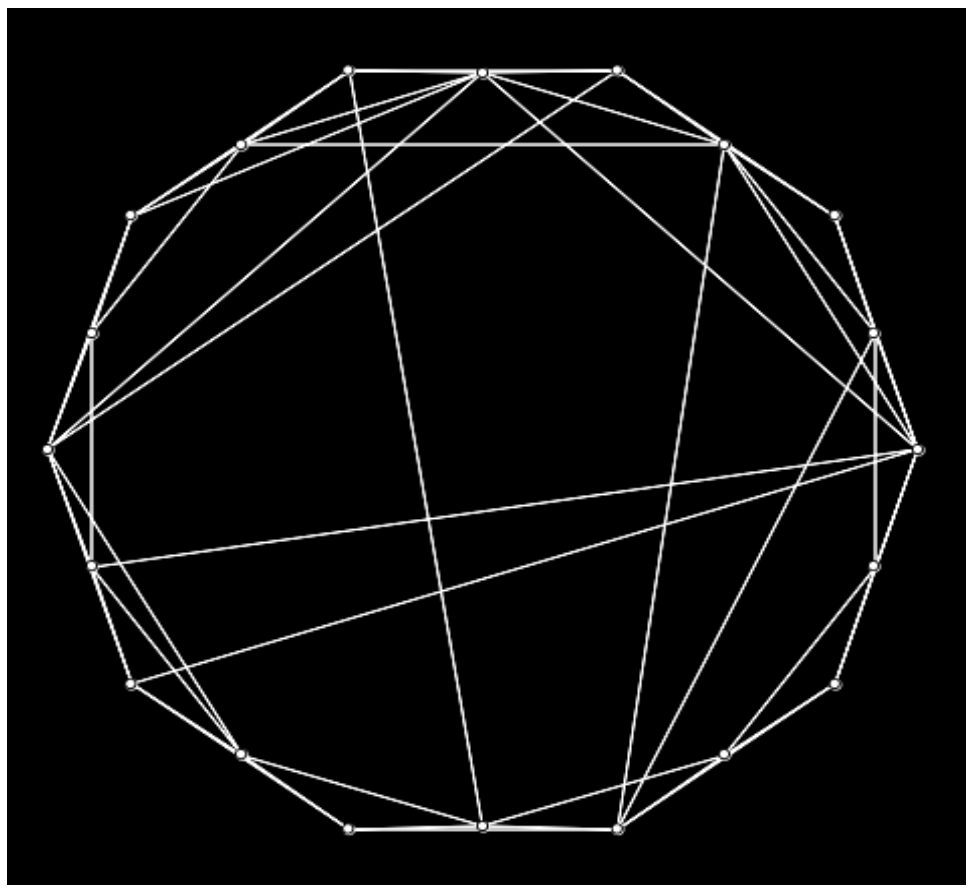

c) Probability = 0.30

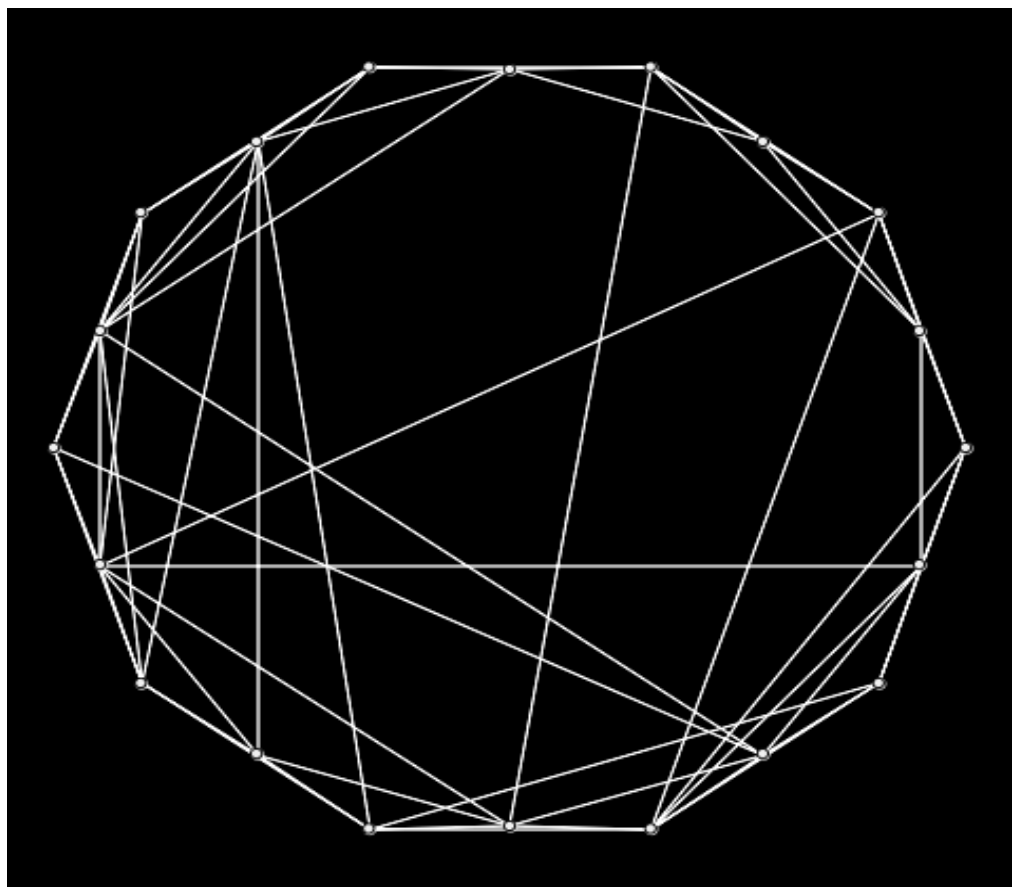

d) Probability =0.50

The simulation results for the average degree of separation for the above graphs are shown in figure 7 below. The average degree of separation is calculated between the nodes as the probability of connectivity increases from *0.00* to *0.50.* From figure 6 it is clear that as the probability increases the number of edges in the graph increases and for a particular value it depicts a social network.

The graph in figure 7 explains how the degree of separation decreases as the probability of connectivity between the nodes increases. Initially the value drops rapidly and later on in the table we observe that the value decreases gradually as the probability increases.

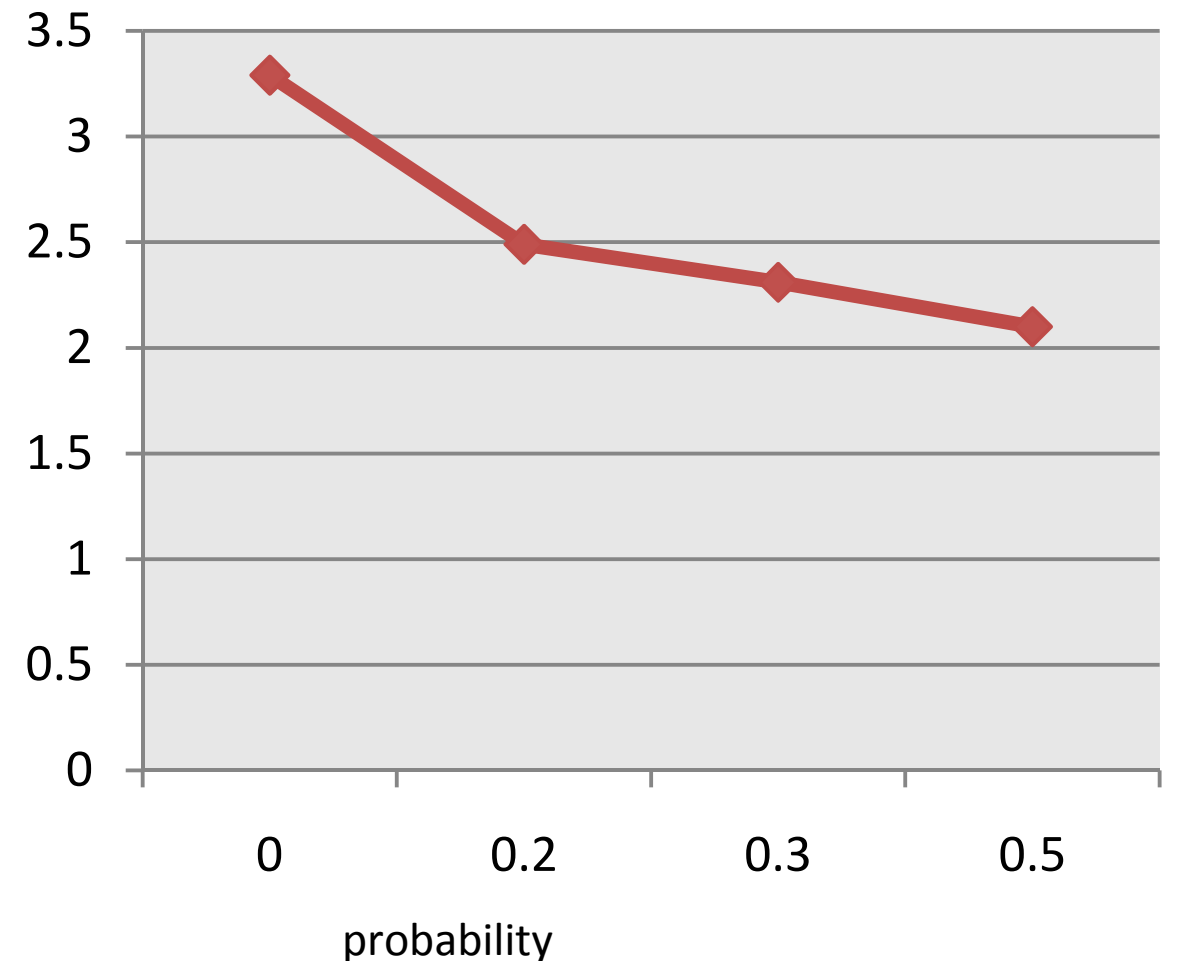

| *Probability* | *Degree of separation* |
|---|---|
| *0.0* | *3.29* |
| *0.2* | *2.49* |
| *0.3* | *2.31* |
| *0.5* | *2.10* |

Figure 7: Graph showing the simulation results of the average degree of separation against the probability of connectivity and the table depicting the values that were obtained during the simulation

## IV CONCLUSION

The paper is motivated by the study of properties of agents in an ecological network and here we consider the question of degree of separation and as a follow we intend to consider other characteristics of ecological networks. We proposed two different strands to find the average degree of separation based on the total number of nodes and the number of links that are associated with each node. In the first method we arrange the given number of nodes a particular sequential tree like structure and then find the degree of separation between them. In the second node we use a matrix to find this value. The second method can be applied for any type of graph. From the various simulation results we have also compared the average degree of separation to the changing probability of connectivity in the graph.